\documentclass[
aps,
nofootinbib,
groupedaddress,
superscriptaddress,
]{revtex4-2}
\input{preamble.sty}

\begin{abstract}
    In this study, we estimate parameters in stochastic oscillatory systems by developing a novel cost function. This function incorporates power spectral density, analytic signal, and position crossings, each weighted to capture distinct oscillatory characteristics such as amplitude, frequency, and shape. By minimizing this cost via differential evolution, we estimate parameters in two stochastic systems given measured datasets. We validate this procedure by recovering known parameters from a test dataset. We then apply it to a biophysical model for auditory mechanics. Thus, we establish a general methodology for fitting stochastic oscillatory systems.
\end{abstract}

\begin{document}
    \title{Estimating Free Parameters in Stochastic Oscillatory Models \\ Using a Weighted Cost Function}
    
    \author{Joseph M. Marcinik}
    \affiliation{Department of Physics \& Astronomy, University of California, Los Angeles, California, 90095, USA}
    
    \author{Dzmitry Vaido}
    \affiliation{Department of Physics \& Astronomy, University of California, Los Angeles, California, 90095, USA}
        
    \author{Dolores Bozovic}
    \affiliation{Department of Physics \& Astronomy, University of California, Los Angeles, California, 90095, USA}
    \affiliation{California NanoSystems Institute, University of California, Los Angeles, California, 90095, USA}
    
    \date{\today}
    \maketitle

    \section{Introduction}
    Scientific models aim to describe natural phenomena with the goal of predicting future outcomes. Achieving accurate predictions hinges on selecting suitable parameter values, which poses significant challenges for models containing many free parameters, such as those encountered in reaction kinetics \cite{mortlockDynamicRegulationJAKSTAT2021,lindenBayesianParameterEstimation2022} and climatology \cite{knightAssociationParameterSoftware2007}. This difficulty further intensifies when models incorporate inherent randomness, as stochastic models must handle not only unavoidable measurement uncertainty but also intrinsic physical noise.
    
    Biological processes naturally exhibit random fluctuations, which typically involve numerous free parameters. These fluctuations typically arise from operating in warm fluid environments. Examples of biological noise sources include thermal fluctuations within organic tissue, shot noise of ion-channel gating, and mechanical clatter of molecular motors. Thus, noisy measurements of biological systems render direct fits from theoretical models computationally impractical, especially when the underlying dynamics exhibit chaos \cite{bakerChaoticDynamicsIntroduction2002,sprottChaosTimeseriesAnalysis2006,faberNoiseinducedChaosSignal2019,faberCriticalityChaosAuditory2024}.
    
    To model noisy biological processes, researchers incorporate stochastic differential equations (SDEs) and Markov processes \cite{gardinerHandbookStochasticMethods2004}. These frameworks have successfully described diverse stochastic phenomena, including smartphone acceleration \cite{sarkkaAdaptiveKalmanFiltering2015}, heartbeats and respiratory cycles \cite{sarkkaDynamicRetrospectiveFiltering2012}, pancreatic $\beta$-cells \cite{cookPancreaticCellsAre1991,bakerChaoticDynamicsIntroduction2002}, and Brownian motion \cite{talebDynamicHedgingManaging1997,liMeasurementInstantaneousVelocity2010}. These stochastic approaches provide powerful means to model complex systems with underlying random processes.
    
    For SDEs, common parameter-estimation methods incorporate Bayesian statistics. For example, linear noise approximation estimates parameter values \cite{zimmerDeterministicInferenceStochastic2015,zimmerReconstructingHiddenStates2015}. This technique, though yielding reliable fits for fast-computing systems, remains computationally intensive and often impractical for slow simulations. As another example, Markov chain Monte Carlo approximates distributions for maximum-likelihood parameters given a dataset, but the corresponding maximum-likelihood models \cite{webbNaiveBayes2017} may undervalue the desired features in the measured system. Hence, these common methods struggle with large stochastic systems because they required either 1) numerically precise likelihood probabilities, which compute too slowly, or 2) analytically intractable likelihood functions, which prove too difficult to derive \cite{cranmerFrontierSimulationbasedInference2020}.
    
    As a concrete application for SDEs, we examine models for the auditory system. In the inner ear, vertebrates rely on hair cells to hear and balance. These sensory cells convert sound waves into neuronal signals with stochastic fluctuations arising from active, nonlinear processes \cite{hudspethMakingEffortListen2008}. For hair cells, there exist two broad categories of models, namely biophysical and phenomenological. Biophysical models incorporate detailed descriptions of experiments performed on biological preparations \cite{martinSpontaneousOscillationHair2003,nadrowskiActiveHairbundleMotility2004,omaoileidighDiverseEffectsMechanical2012,marcinikComparingParameterreductionMethods2024}. In contrast, phenomenological models apply nonlinear dynamics and bifurcation theory to describe biophysical phenomena \cite{hudspethCritiqueCriticalCochlea2010,faberChaoticDynamicsEnhance2019,hudspethCriticalThingEars2024,metzler-winslowNeuralControlInnate2024}.
    
    Biophysical models generally incorporate numerous free parameters. These parameters account for many cellular processes, capturing the intricate dynamics of auditory hair cells. In auditory models, free parameter values vary across experiments, depending on type of hair cell \cite{marcottiHairCells2010}, type of end organ \cite{smothermanHearingFrogs2000}, and species of host animal (e.g., chick \cite{ohmoriMechanoelectricalTransductionCurrents1985}, turtle \cite{crawfordMechanicalPropertiesCiliary1985}, frog \cite{martinSpontaneousOscillationHair2003}).
    
    Although some free parameters appear constrained by experiments, many remain unmeasured. To constrain these parameter values experimentally for individual cells, we must overcome various experimental obstacles, such as measuring long recording times, dissecting fragile biological tissue, and observing elusive internal processes. Thus, we instead estimate free parameter values numerically to model specific cells. However, simulating stochastic time series involves burdensome computations.
    
    To address this challenge, we need robust, compute-friendly techniques to estimate parameters. Prior studies have applied simulation-based inference \cite{cranmerFrontierSimulationbasedInference2020} to simpler phenomenological models for hair cells, such as the Duffing equation \cite{belousovVolterraseriesApproachStochastic2019}, the van der Pol oscillator \cite{belousovVolterraseriesApproachStochastic2020}, a hybrid of these two \cite{thipmaungpromActiveEnergyHarvesting2025}, and a generalized Ornstein-Uhlenbeck process \cite{tucciModelingActiveNonMarkovian2022}. In this study, we focus on biophysical hair-cell models, which contain high-dimensional systems of coupled SDEs and numerous free parameters. These high-dimensional biophysical models generally require more computing power than their low-dimensional phenomenological counterparts. Furthermore, they benefit from fits to longer datasets, which requires more computing power.

    To bypass directly fitting time series, modelers define cost functions. This function, as input, takes a specific instance of free parameters. As its output, it returns a cost---a scalar that represents deviation between the simulation and the measured dataset. By minimizing this cost, modelers can optimize their free parameters to match their desired features, such as amplitude or frequency. Unfortunately, crafting an effective cost function generally proves nontrivial and usually requires system-specific tailoring, especially for stochastic systems.
    
    Accordingly, we develop a fast numerical procedure to fit prolonged noisy measurements to high-dimensional stochastic models. To optimize parameter values, we establish a cost function, tailoring it toward oscillatory SDEs. We evaluate this procedure by estimating parameters given long recordings of hair-bundle motion, demonstrating its fruitfulness for biological data.
    
    We organize this paper as follows. In \cref{sec:methods}, we overview the numerical optimization for the cost function, and we outline the \textit{ex vivo} conditions for the hair cells. In \cref{sec:derivation}, we describe the biophysical system of SDEs for hair cells, reproducing the spontaneous oscillations observed in auditory mechanics. In \cref{sec:cost}, we delineate the weighted cost function, adjusted to optimize parameter fits for the biophysical model. In \cref{sec:cc}, we compare the simulated models to their measured counterparts by calculating their cross-correlation. In \cref{sec:discussion}, we analyze and summarize the findings obtained from the biophysical model. We also discuss how to interpret the cost function and extend it to other systems.

    \section{Methods} \label{sec:methods}
    We devised methods to optimize parameter values in a stochastic model for active hair-bundle oscillations. We applied it to experimental datasets obtained from live hair cells, leveraging a cost function to evaluate the mismatch between measured and simulated traces of motion.

    \subsection{Applying Differential Evolution}
    We minimized the cost function numerically using differential evolution \cite{stornDifferentialEvolutionSimple1997,priceDifferentialEvolution2005}. This algorithm has received wide usage in a variety of fields, including biology \cite{oguzOptimizationModelReduction2013,defalco5DifferentialEvolution2021} and physics \cite{hellerDifferentialEvolutionOptimization2022,yuanImprovedDifferentialEvolution2023,aljaidiTwoStageDifferential2025}. Applying differential evolution, we optimized parameter values for individual hair bundles. During each fit, we performed 2000 iterations within 12 hours using an Intel i9-13900K (24 cores, 32 threads). For more details on differential evolution, refer to \cref{app:de}.

    \subsection{Rescaling the Nondimensional Simulation} \label{sec:rescale}
    In an earlier study, we showed that nondimensionalizing a deterministic model (for hair-bundle mechanics) naturally reduced its number of free parameters \cite{marcinikComparingParameterreductionMethods2024}. After optimizing parameter values, this noiseless model reproduced individual measured oscillations of hair-bundle motion. Extending this approach, we assimilated a stochastic version of this model to fit multiple oscillation cycles simultaneously. By including stochastic terms, we captured variations in instantaneous frequency and amplitude as well as oscillation shape, observed experimentally.
    
    To compare simulations with measurements, we rescaled from the simulated nondimensional units into the measured dimensional units. Specifically, we fit three rescaling factors $\chr{x}_{hb}$, $\check{x}_{hb}$, and $\chr{\tau}$ defined by
    \begin{align} \label{eq:rescale}
        X_{hb}(t) = \chr{x}_{hb} \qty(\nondim{x}_{hb}(\nondim{t}) - \check{x}_{hb})
        && \nondim{t} = \frac{t}{\chr{\tau}}
    \end{align}
    where $\chr{x}_{hb}$ and $\check{x}_{hb}$ represent multiplicative scaling and constant offset, respectively, of nondimensional bundle position $\nondim{x}_{hb}$; and $\chr{\tau}$ represents multiplicative scaling for nondimensional time $\nondim{t}$. To reduce computing time, we independently rescaled each component of the cost function instead of directly rescaling the position, as described in \cref{app:cost-function}.

    \subsection{Defining our Mathematical Notation}
    Throughout this study, we employ the following mathematical notation for brevity. We represent the number of occurrences in which a random variable satisfies the event or condition $X$ as $N[X]$. We denote expected value of quantity $x$ as $\expval{x}$. We indicate nondimensional or characteristic quantities with the tilde ($\,\nondim{\cdot}\,$) or hat ($\hat{\cdot}$) accents, respectively.
    
    To describe collections of mathematical objects, we often use tuples with consistent notation. We describe our notation for tuples. We denote the $n^{th}$ element of a tuple $S$ as $(S)_n$, dropping the parentheses implicitly as $S_n$ when appropriate. We denote the tuple $(S)_\mathfrak{N}\coloneq\qty(S_n|n\in\mathfrak{N})$ given a set $\mathfrak{N}$ of indices, similarly dropping the parentheses as $S_\mathfrak{N}$ when appropriate.
    
    To distinguish between the two types of normalization performed in this paper, we motivate our notation from Lebesgue spaces $L^n$ \cite{gowersMathematicalConcepts2008}. We $L^1$-normalize a function $F(x)$ by taking $F(x)\mapsto F(x)/\int\abs{F(x)}\dd{x}$. We $L^\infty$-normalize a function by taking $F(x)\mapsto F(x)/\max_x{\abs{F(x)}}$, and we deem a function that ranges between 0 and 1, inclusively, as $L^\infty$-normalized. We note that $L^\infty$-normalized functions, in general, range between -1 and 1, inclusively; however, we $L^\infty$-normalize only non-negative functions in this paper, thus leading to a range between 0 and 1, inclusively.

    \subsection{Preparing the Biological Experiment}
    We used the North American bullfrog (\textit{Rana catesbeiana}) as our biological system. Its sacculus---a hearing end organ responsible for detecting vestibular movement and low-frequency sounds---has been widely used in hearing experiments due to its robust structure, optical accessibility, and physiological similarity to mammalian auditory systems \cite{howardMechanicalRelaxationHair1987,hudspethMakingEffortListen2008}. After dissecting a sacculus, we ensured that it maintained sufficient physiological integrity.
    
    From dissected sacculi, we imaged hair cells \textit{ex vivo}. We optically tracked the motion of hair bundles---organelles comprised of roughly 50 stereocilia protruding from the apical surface of hair cells \cite{benserRapidActiveHair1996}. These motion measurements produced traces of active hair-bundle motion, as illustrated in \cref{fig:crossing-demo,fig:cc-trace}, helping to validate numerical simulations against experimental observations. From these recordings, we manually selected nine spontaneously oscillating bundles.
    
    To assess our parameter fits, we compared parameter values obtained from saccular data to those from amphibian papilla (AP) data. The AP---a hearing end organ responsible for detecting high-frequency sounds---contains hair bundles that oscillate spontaneously. Morphologically, these hair bundles appear distinct to those in the sacculus \cite{smothermanHearingFrogs2000}, making the AP an ideal candidate for contrasting to the sacculus. We dissected AP tissue following our previous work \cite{vaidoSpontaneousOscillationsHair2025} and recorded cells following the aforementioned imaging techniques. From these recordings, we manually selected ten spontaneously oscillating bundles. For more details on our biological preparation and optical recordings, refer to \cref{app:methods}.
    
    \section{Theoretical Model of Spontaneous Hair-Bundle Motility} \label{sec:derivation}
    To model the spontaneous oscillations observed in inner-ear hair cells, we derived a system of SDEs as shown in \cref{tbl:nondim}. These oscillations, driven by various cellular processes, manifest as active movements of the stereocilia---a bundle of strand-like sensory organelles atop each hair cell. Stereocilia contain mechano-sensitive ion channels and associated molecular machinery, both essential for translating sound waves into neuronal signals. They oscillate spontaneously, according to previous experiments conducted \textit{in vitro} and \textit{ex vivo} \cite{hudspethMakingEffortListen2008,hudspethIntegratingActiveProcess2014}.
    
    These oscillations originate from three key molecular processes, namely mechano-electrical transduction (MET; \cite{ohmoriMechanoelectricalTransductionCurrents1985,eatockAdaptationMechanoelectricalTransduction1987,howardMechanoelectricalTransductionHair1988}), myosin-mediated adaptation \cite{walkerCalmodulinCalmodulinbindingProteins1993,walkerCalmodulinControlsAdaptation1996,cyrMyosin1cInteractsHairCell2002,holtChemicalGeneticStrategyImplicates2002,gillespieMyosin1cHairCells2004}, and tip-link tension \cite{hacohenRegulationTensionHaircell1989,assadTiplinkIntegrityMechanical1991,martinNegativeHairbundleStiffness2000}. Together with the ion dynamics that modulate them, these processes form the foundation of the proposed biophysical model \cite{martinSpontaneousOscillationHair2003}.
    
    In our previous study, we explored a deterministic model of hair-bundle mechanics, gleaning key experimental insights from hair-cell literature \cite{marcinikComparingParameterreductionMethods2024}. We first systematically reduced this comprehensive biophysical model to a 15-parameter nondimensional model, optimizing it for fitting without loss of generality. We subsequently fixed the least influential parameters sequentially, identifying a five-parameter deterministic model that maximizes predictive power when fitting an individual cycle. We hence concluded that this five-parameter model optimally captured the features of regular hair-bundle oscillations.
    
    Deterministic models, however, fail to fully capture the measured variation in hair-bundle movement. Most recordings display substantial fluctuations in frequency, amplitude, and shape of oscillations, necessitating a stochastic approach. To emulate this noise, we adopted a stochastic model capable of fitting multiple cycles simultaneously, rather than a deterministic model capable of fitting single cycles individually.
    
    Our stochastic model employs an eight-parameter system of SDEs, with three degrees of freedom for rescaling, adapting our previous five-parameter deterministic system of ODEs \cite{marcinikComparingParameterreductionMethods2024}. This model incorporates two noise terms $\nondim{\eta}_{hb}$ and $\nondim{\eta}_x$, which represent fluctuations in hair-bundle position $\nondim{x}_{hb}$ and adaptation-motor position $\nondim{x}_a$, respectively. The full system used to simulate hair-bundle motion is summarized in \cref{tbl:nondim}, with analytic derivations and computational methods detailed in \cref{app:model}.
    
    \begin{table*}
        \small
        \begin{tabular*}{\textwidth}{llll}
            \toprule\midrule
            Quantity & Significance & Equation & Type \\\midrule
            $\nondim{x}_{hb}$ & Hair-bundle position & $\nondim{\tau}_{hb} \dot{\nondim{x}}_{hb} = -\qty(\nondim{F}_{gs}+\nondim{x}_{hb})+\nondim{\eta}_{hb}\dot{\nondim{B}}_{\nondim{t}}$ & Variable \\
            $\nondim{x}_a$ & Myosin-motor position & $\dot{\nondim{x}}_a = \nondim{S}_{max}\nondim{S}\qty(\nondim{F}_{gs}-\nondim{x}_a)-\qty(1-\nondim{S}_{max})+\nondim{\eta}_a\dot{\nondim{B}}_{\nondim{t}}$ & \\
            $p_m$ & $\text{Ca}^{2+}$-binding probability for myosin motor & $\nondim{\tau}_m \dot{p}_m = p_T(1-p_m)-p_m$ & \\
            $p_{gs}$ & $\text{Ca}^{2+}$-binding probability for gating spring & $\dot{p}_{gs} = \nondim{C}_{gs}p_T(1-p_{gs})-p_{gs}$ & \\\midrule
            $\nondim{k}_{gs}$ & Dynamic stiffness for gating spring & $\nondim{k}_{gs} = 1-p_{gs}\qty(1-\nondim{k}_{gs,min})$ & Function \\
            $\nondim{F}_{gs}$ & Effective force (gating/extent springs, pivot) & $\nondim{F}_{gs} = \nondim{k}_{gs}\qty(\nondim{x}_{hb}-\nondim{\chi}_a\nondim{x}_a-p_T)$ & \\
            $\nondim{S}$ & Slipping rate for myosin motor & $\nondim{S}=-p_m\qty(1-\nondim{S}_{min})$ & \\
            $p_T$ & Open probability for transduction channel & $p_T=\qty[1+\exp(\nondim{U}_{gs,max}\qty(\Delta{\nondim{E}}^\varnothing-\nondim{k}_{gs}\qty(\nondim{x}_{gs}-\frac{1}{2})))]^{-1}$ & \\
            \midrule\bottomrule
        \end{tabular*}
        \caption{Variables and functions in the stochastic nondimensional model for hair-bundle mechanics. This model consists of eight deterministic parameters, two noise strengths, and three rescaling degrees of freedom. For each quantity, the \enquote{Significance} column indicates its physical significance, and the \enquote{Equation} column shows its defining equation. The \enquote{Type} column indicates whether the quantity represents a variable or function. The dot derivative indicates a derivative with respect to nondimensional time $\nondim{t}$.} \label{tbl:nondim}
    \end{table*}

    \section{Weighted Components of Cost Function} \label{sec:cost}	
    We defined a cost function with three components: power spectral density, analytic signal, and position crossings, all illustrated in \cref{fig:cost-cell,fig:cost-demo}. For each component, we formulated an $L^\infty$-normalized cost to quantify the dissimilarity between measured and simulated traces. By $L^\infty$-normalizing, we assigned comparable costs to each component, which allowed us to emphasize the desired properties of numerical simulations relative to measured recordings.
    
    To quantify dissimilarity between $L^1$-normalized functions, we employed the total variation distance (TVD)---an information-theoretic measure of $L^\infty$-normalized distance \cite{tsybakovIntroductionNonparametricEstimation2009} between PDFs. To calculate the TVD, we computed
    \begin{equation}
        d_\text{TV} \qty(\rho_1, \rho_2)
        = \frac{1}{2} \int \abs{\rho_1(x) - \rho_2(x)} \dd{x}
    \end{equation}
    where $\rho_1$ and $\rho_2$ represent two probability density functions (PDFs). Conceptually, a larger TVD indicates a greater dissimilarity between distributions. Throughout this work, we use TVD to quantify the dissimilarity between $L^1$-normalized functions, such as PDFs.
    
    To denote the tuple of $L^\infty$-normalized costs, one for each component in the cost function, we used $vec{C}$. After defining an $L^1$-normalized weight vector $\vec{w}=(0.1,0.5,0.4)$, we asserted $\vec{w}\cdot\vec{C}$ as the overall cost between two traces. Similar to its constituent costs, this weighted cost remains $L^\infty$-normalized. By adopting a \textit{weighted} cost function, we emphasized desired features of the numerical fits relative to measured datasets. Weighted cost functions have proven effective across diverse fields, including healthcare \cite{parrSolvingMultiobjectiveNurse2007,veerkampEvaluatingCostFunction2021}, economics \cite{adlerBenefitCostAnalysis2016,miettinenCostFunctionApproach2020,alviOptimalityConditionDistributed2022}, and mathematics \cite{wuWeightedlossfunctionControlCharts2006,hoRealWorldWeightCrossEntropyLoss2020,bandyDynamicWeightingFactor2024,faberCriticalityChaosAuditory2024}. For more details on the definition and computation of this cost function, refer to \cref{app:cost-function}.

    \subsection{Shape of Power Spectral Density} \label{sec:psd}
    Let $\widetilde{S}_{xx}\qty{x}(f)$ represent the $L^1$-normalized power spectral density of position $x(t)$. In the measured recordings, this position denotes the time-dependent trace of hair-bundle position, observed optically. The power spectral density characterizes the frequency distribution of a trace, offering insight into its spectral composition.
    
    To quantify the discrepancy between $\widetilde{S}_{xx}\qty{x}$ from the measured and simulated traces, we computed their TVD. Here, a larger TVD indicates a greater difference in spectral content between the two traces. For computational details on this power spectral density, refer to \cref{app:psd}.

    \subsection{Distribution of Analytic Signal} \label{sec:das}
    Let $S_a\{x\}(t)=x(t)+i\mathcal{H}\qty{x}(t)$ represent the analytic signal of position $x(t)$, where $\mathcal{H}$ denotes the Hilbert transform. The analytic signal encodes information about the amplitude and phase distributions for an oscillation \cite{vakmanAmplitudePhaseFrequency1977,justiceAnalyticSignalProcessing1979,vakmanAnalyticSignalTeagerKaiser1996}. It also holds statistics for the distribution of noise for a trace. To represent the distribution of the analytic signal \cite{feldmanHilbertTransformVibration2011,salviIdentificationBifurcationsObservations2016}, consider the 2D joint PDF
    \begin{equation} \label{eq:asigp}
        \rho_a(x,\mathcal{H}\qty{x}) \propto N\qty[\nasig{x}=\widetilde{s}_a]
    \end{equation}
    of the $L^1$-normalized analytic signal $\nasig{x}$.
    
    To quantify the discrepancy between $\nasig{x}$ from the measured and simulated traces, we computed their TVD. Here, a larger TVD indicates a greater dissimilarity in amplitude, phase, or noise between the two traces. For computational details on this distribution of the analytic signal, refer to \cref{app:das}.

    \subsection{Distribution of Position Crossings} \label{sec:dpc}
    Let $\zc{x}$ represent the ordered tuple of zero-crossing times from position $x(t)$, sorted in ascending order of time. These zero-crossings encode information primarily about the frequency and shape distributions for an oscillation \cite{kayZeroCrossingbasedSpectrum1986,hsueAutomaticModulationClassification1990,sekharAdaptiveWindowZeroCrossingBased2004,ramunno-johnsonDistributionFrequenciesSpontaneous2009}. For a general crossing level $\gamma\in\mathbb{R}$, define the ordered tuple of $\gamma$-crossing times as $\zc[\gamma]{x}\coloneq\zc[0]{x-\gamma}$. To characterize the interval between consecutive $\gamma$-crossing times in $\zc[\gamma]{x}$, we defined the unordered tuple $\Delta\zc[\gamma]{x}$ by
    \begin{equation}
        \qty(\Delta\zc[\gamma]{x})_{n+1}
        \coloneq \qty(\zc[\gamma]{x})_{n+1} - \qty(\zc[\gamma]{x})_n
    \end{equation}
    for $n\in\mathbb{N}$. Conceptually, the tuple $\Delta\zc[\gamma]{x}$ resembles half-periods of an oscillation.
    
\begin{figure}
        \centering
        \includegraphics{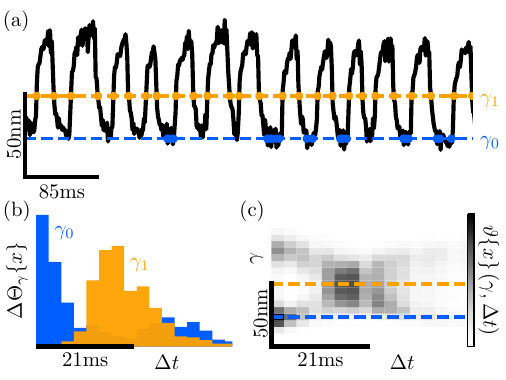}
        \caption{A walkthrough for the distribution of position crossings, as described in \cref{sec:dpc}. Blue and orange colors correspond to $\gamma$-crossings at $\gamma=\gamma_0$ and $\gamma=\gamma_1$, respectively. (a) The solid line displays position $x$ over time $t$ for a measured hair bundle. The two dashed lines indicate where $x=\gamma_0$ (blue) or $\gamma_1$ (orange), and markers correspond to times of $\gamma$-crossing events. The scale bar displays one amplitude ($\SI{50}{\nano\meter}$; half peak-to-peak) and two periods ($\SI{85}{\milli\second}$) of oscillation. (b) The two histograms show PDF $\Delta\zc[\gamma]{x}$ over time difference $\Delta{t}$ for $\gamma=\gamma_0$ (blue) or $\gamma_1$ (orange). The scale bar displays half-period ($\SI{21}{\milli\second}$). (c) The heatmap indicates PDF $\vartheta{x}(\gamma,\Delta{t})$ over crossing position $\gamma$ and time difference, ranging from zero (white) to maximum (black). The two dashed lines indicate where $\gamma=\gamma_0$ (blue) or $\gamma_1$ (orange). The scale bar displays amplitude and half-period ($\SI{21}{\milli\second}$).} \label{fig:crossing-demo}
    \end{figure}

    We defined the distribution of position crossings as the 2D joint PDF
    \begin{equation}
        \vartheta\qty{x}(\gamma,\Delta{t})
        \propto N \qty[x=\gamma \land \Delta\zc[\gamma]{x}=\Delta{t}]
    \end{equation}
    where $N$ denotes the number of $\gamma$-crossings with half-periods $\Delta{t}$. To quantify the discrepancy between two $\vartheta\qty{x}$ from the measured and simulated traces, we computed their TVD. A larger TVD indicates a greater dissimilarity in frequency or shape between two traces. For computational details on this distribution of position crossings, refer to \cref{app:dpc,fig:crossing-demo}.
    
    \subsection{Validation on Triangle Waves} \label{sec:triangle-recovery}
    To evaluate our cost function, we tested it on a controlled dataset of triangle waves. We generated waves with the form
    \begin{equation} \label{eq:triangle}
        \triangle(t,w)
        = A \nondim{\triangle}\qty(2\pi ft, w) + x_0
        + \mathcal{N}\qty(0,\sigma_\triangle^2)
    \end{equation}
    where $\mathcal{N}\qty(0,\sigma^2)$ represents a normal random variable with zero mean and variance $\sigma^2$. Here, the five parameters represent amplitude $A$, frequency $f$, offset $x_0$, width $w$, and noise strength $\sigma_\triangle$.
    
    We simulated 600 triangle waves. For each simulation, we varied all five parameters simultaneously within a predefined range, ensuring a diversity of waveforms. By applying differential evolution to minimize the cost function, we recovered all 600 sets of known parameters. These results demonstrate that the cost function can successfully recover optimal parameter sets, as shown in \cref{fig:triangle-recovery}, even in the presence of noise. For more details on the computational methods and parameter ranges, refer to \cref{app:triangle-recovery}.

    \begin{figure*}
        \centering
        \includegraphics{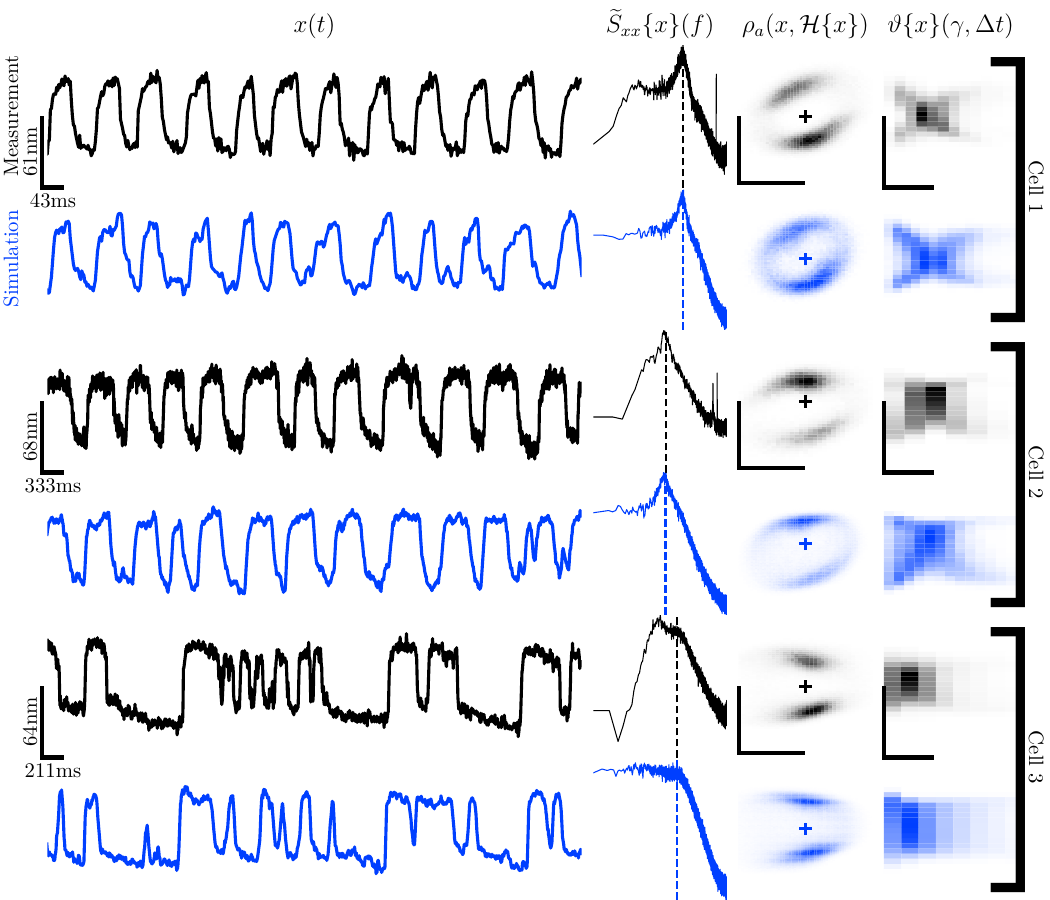}
        \caption{The three cost-function components, described in \cref{sec:cost}, for three saccular bundle pairs. Black and blue colors indicate measured and simulated hair bundles, respectively. Each pair of rows corresponds to a measured bundle (black) and its concomitant simulation (blue). Column 1: The solid lines display position $x$ over time $t$ for measured and simulated hair bundles, each with zero mean position. The scale bars display amplitudes of $2\langle\vert\nasig{x}\vert\rangle$ along the $y$-axis and timescales of half-period along the $x$-axis, shared between each pair of measured and simulated traces. Column 2: In a log-log scale, the solid lines display power spectral density $\widetilde{S}_{xx}$ over frequency $f$, and the dashed lines indicate their median frequency. Column 3: The heatmaps display PDF $\rho_a(x,\mathcal{H}\qty{x})$ over position along the $y$-axis and Hilbert transform $\mathcal{H}\qty{x}$ along the $x$-axis. The plus signs indicate the origin of $(x,\mathcal{H}\qty{x})$. The scale bars display amplitude along both axes. Column 4: The heatmaps display PDF $\vartheta\{x\}(\gamma,\Delta{t})$ over crossing position $\gamma$ along the $y$-axis and time difference $\Delta{t}$ along the $x$-axis. The scale bars display one amplitude along the $y$-axis and half period along the $x$-axis.} \label{fig:cost-cell}
    \end{figure*}

    \section{Quantified Similarity Between Two Traces} \label{sec:cc}
    \subsection{Distribution of Cross-Correlations} \label{sec:cc-dist}
    To quantify the similarity between two traces, we developed a method to compare time series. Here, we consider two traces $x(t)$ and $y(t)$, possibly with two identical traces.
    
    First, we rescaled $y$ to match $x$ by optimizing the three rescaling variables in \cref{sec:rescale} using the cost function described in \cref{sec:cost}. We then partitioned $x$ and $y$ into unordered tuples of equal-duration segments $x_\mathbb{N}$ and $y_\mathbb{N}$, respectively, each of duration $T$. By comparing these segments, we captured not only the expected match between traces but also their plausible range of agreement.
    
    Next, we considered each pair of segments in $\zeta_T\qty{x,y}\coloneq x_\mathbb{N}\times y_\mathbb{N}$, i.e., one segment each from $x$ and $y$. For each pair $(x_n,y_m)\in\zeta_T\qty{x,y}$, we calculated their cross-correlation $\hat{C}_{nm}(\Delta{t})\propto\expval{x_n(t)y_m(t+\Delta{t})}$ over a time lag $\Delta{t}$. We then $L^\infty$-normalized this cross-correlation to ensure reliable comparisons across different pairs of traces.
    
    Finally, we defined the cross-correlation matrix as $\vec{C}_{max}\qty{x,y}$, in which the $(n,m)^{th}$ element equals $\max_{\Delta{t}}\hat{C}_{nm}(\Delta{t})$ for all $n,m\in\mathbb{N}$. Conceptually, each element represents the maximum cross-correlation between $x_n$ and $y_m$ over all time lags. We further defined the PDF for the flattened matrix $\vec{C}_{max}$ as $\chi\qty{x,y}(\hat{C})$, where $\hat{C}$ denotes cross-correlation values. This distribution quantifies the overall correlation between traces $x$ and $y$, as illustrated in \cref{fig:cc-trace}.

    \subsection{Reduction to Jensen-Shannon Divergence}
    Let $x(t)$ represent a reference trace and $y(t)$ a comparison trace. To quantify their dissimilarity, we calculated the Jensen-Shannon divergence $\jsd{x}{y}$, an information-theoretic measure of $L^\infty$-normalized distance between two probability distributions \cite{polyanskiyInformationTheoryCoding2022}. Conceptually, this divergence measures information loss when two distributions try to approximate each other. We prefer it over the related Kullback-Leibler divergence because the latter remains unbounded, complicating comparisons across different datasets. For comparisons to other information-theoretic measures, refer to \cref{app:measure-compare}.

    \subsection{Matrix of Jensen-Shannon Divergences} \label{sec:jsd-matrix}
    Let $x(t)$ and $y(t)$ continue to represent reference and comparison traces, respectively. Our measured traces (typically reference traces) consist of experimental recordings of hair-bundle position, while our simulated traces (typically comparison traces) come from numerical simulations of bundle position.
    
\begin{figure}
        \centering
        \includegraphics{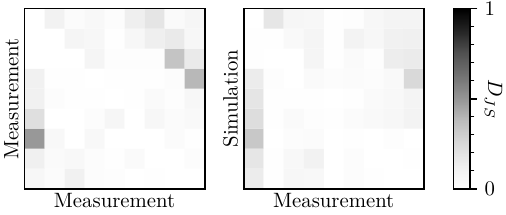}
        \caption{Two Jensen-Shannon divergence matrices, as described in \cref{sec:jsd-matrix}, for nine saccular bundles. The colorbar indicates values of Jensen-Shannon divergences between reference and comparison bundles. Left: The heatmap displays a measured-measured divergence matrix comparing nine saccular bundles. The $x$-axis indicates index of measured reference bundle, and the $y$-axis indicates index of measured comparison bundle. Right: The heatmap displays a measured-simulated divergence matrix. The $x$-axis indicates index of measured reference bundle, and the $y$-axis indicates index of simulated comparison bundle.} \label{fig:jsd}
    \end{figure}

    We computed $D_{JS}$ for each pair of traces in two groups: 1) measured reference with measured comparison traces and 2) measured reference with simulated comparison traces, with the latter illustrated in \cref{fig:cc-distribution}. Displayed in \cref{fig:jsd,fig:divergence-compare}, these comparisons produced two asymmetric divergence matrices of $D_{JS}$, which we call the \enquote{measured-measured} and \enquote{measured-simulated} divergence matrices, respectively. Conceptually, a measured-measured matrix reflects the degree of variation among measured traces, while a measured-simulated matrix reflects the level of disagreement between measured and simulated traces.
    
    \begin{figure}
        \centering
        \includegraphics{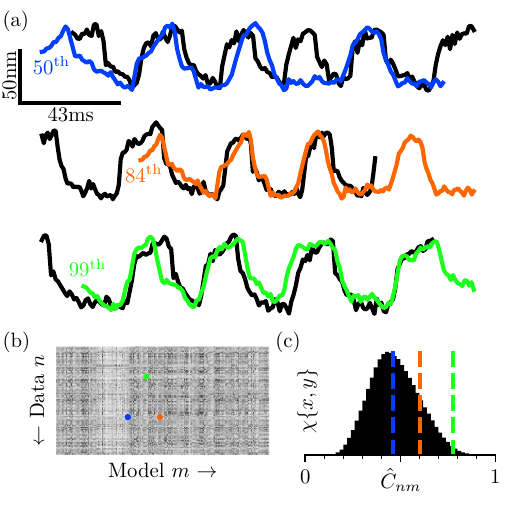}
        \caption{A walkthrough for the matrix of cross-correlations described in \cref{sec:cc-dist}. The three colors correspond to different percentiles of $\chi\qty{x,y}$: 50$^{\text{th}}$ in blue; 84$^{\text{th}}$, orange; 99$^{\text{th}}$, green. (a) The lines display measured traces $x$ (black) and simulated traces $y$ (colored). The colors indicate measured-simulated segment pairs corresponding to their respective percentiles. (b) The heatmap displays elements in matrix $\vec{C}_{max}$. The heatmap ranges from $\hat{C}=0$ (white) to $\hat{C}=1$ (black). The $y$-axis indicates the index $n$ for the measured segment, increasing in time downward. The $x$-axis indicates the index $m$ for the simulated segment, increasing in time rightward. The colored dots indicate the segment indices corresponding to their respective percentiles. (c) The histogram displays PDF $\chi\qty{x,y}$. The colored dashed lines indicate cross-correlation $\hat{C}_{nm}$ corresponding to their respective percentiles.} \label{fig:cc-trace}
    \end{figure}

    \section{Discussion} \label{sec:discussion}
    \subsection{Benefits of the Cost Function}
    We developed a cost function to fit parameters in oscillatory system of SDEs, prioritizing speed, interpretability, and adaptability.

    \subsubsection{Computational Speed}
    The cost function computes efficiently. The rate-limiting steps in our fitting procedure involve 1) propagating the SDEs forward in time and 2) computing the cost-function components. When rescaling a trace directly using \cref{eq:rescale}, we must repeatedly recalculate the cost function, which requires extensive computational time. Thus, we instead rescaled the cost-function components directly, detailed in \cref{app:cost-function}, providing a quicker rescaling option.

    \subsubsection{Scientific Interpretability}
    The cost function retains scientific interpretability. It remains transparent throughout the fitting procedure, allowing users to conceptualize why particular costs yield certain fit qualities. Specifically, it contains three components, each capturing key aspects of active oscillations. Given a trace $x(t)$, the power spectral density $\widetilde{S}_{xx}\{x\}$ and distribution $\vartheta\qty{x}(\gamma,\Delta{t})$ hold frequency and shape information, while the distribution $\rho_a\qty{x}$ encodes amplitude and phase information. The two distributions had heavier weights because they contained the most relevant details for the hair-bundle oscillations. We still included the power spectral density because it converts timescales economically from nondimensional to dimensional units.

    \subsubsection{General Adaptability}
    Beyond remaining interpretable, the cost function generally fits versatile types of oscillatory dynamics. We, for instance, used the cost function to fit two types of distinct oscillatory behavior, as discussed in \cref{sec:bundle-params}. Researchers can tailor it to their applications, for example, by adjusting component weights, assimilating additional components, or deriving system-specific models.
    
    The cost function accommodates adaptable weights for its components. They can adjust these weights to align with their scientific systems and objectives. To further customize these weights, they can update them dynamically during differential evolution \cite{parrSolvingMultiobjectiveNurse2007,alviOptimalityConditionDistributed2022,bandyDynamicWeightingFactor2024}.
    
    In addition to offering adjustable weights, the cost function can assimilate additional components. For example, we explored incorporating the distribution of velocities from the analytic signal in \cref{app:velocity-field}. We excluded this fourth component due to its sensitivity to noise, but it may benefit systems with higher signal-to-noise ratios.
    
    After including extra components, the cost function may extend to chaotic systems. Prior studies have leveraged concepts from topological data analysis, such as persistence homologies \cite{tymochkoUsingPersistentHomology2020}, CROCKER plots \cite{topazTopologicalDataAnalysis2015,ulmerTopologicalApproachSelecting2019,guzelDetectingBifurcationsDynamical2022}, and Takens's embeddings \cite{takensDetectingStrangeAttractors1981,eckmannErgodicTheoryChaos1985,faberChaoticDynamicsEnhance2019,myersDelayParameterSelection2024}. These approaches could replace the distribution of the analytic signal described in \cref{sec:das}, acting as robust metrics for fitting chaotic systems.
    
    While we estimated parameters in a model for hair-bundle mechanics, our methodology extends to various oscillatory systems. In the current study, we modeled hair-bundle motion in two hearing end organs, the sacculus and AP \cite{vaidoSpontaneousOscillationsHair2025}. To model other inner-ear end organs (e.g., utricle or basilar papilla \cite{smothermanHearingFrogs2000}), researchers can incorporate additional terms or extend parameter ranges before evaluating the cost function. For general oscillatory systems, they must develop suitable models, and they can continue to implement our cost-function approach to characterize and optimize the match between their measured and simulated datasets.

    \subsection{Interpretation of the Divergence Matrix}
    \subsubsection{Distribution of Cross-Correlations}
    The cross-correlation distribution, defined in \cref{sec:cc-dist}, compares \textit{every} pair of segments from two traces, including non-coincident pairs. This approach assumes that the biophysical system exhibits Markovian dynamics, implying that it lacks an absolute time reference. Consequently, we can justifiably compare non-coincident pairs of segments. We emphasize that the median cross-correlation (shown in \cref{fig:cc-trace}) reflects the expected agreement when selecting \textit{any} random pair of segments, one from each trace.

    The cross-correlation distribution requires significantly longer traces for non-Markovian than Markovian systems. Non-Markovian systems yield at most $\Theta\qty(N)$ cross-correlations for this distribution. In contrast, Markovian systems generate up to $\Theta\qty(N^2)$ cross-correlations, offering greater sample sizes and higher numerical precision for estimating the distribution.

    Using the cross-correlation distribution, we analyzed intracellular variation within a measured trace. For a given trace $x(t)$, the distribution $\chi\qty{x,x}$ captures this variation, which arises in hair bundles due to various stochastic processes, as described in \cref{app:model}. To quantify mechanical variation within a single trace, we used $\expval{\chi\qty{x,x}}$. We note that a large $\expval{\chi\qty{x,x}}$ indicates regular oscillations (i.e., with high quality-factor), whereas a small value indicates irregular movement. This value varies with segment duration $T$, so its nominal value communicates minimal information. However, comparing $\expval{\chi\qty{x,x}}$ across different traces elucidates which bundles exhibit more regular oscillations relative to others.

    \subsubsection{Divergence of Cross-Correlations}
    We evaluated the agreement between measured and simulated traces. The simulated traces qualitatively reproduced the measured traces, as shown in \cref{fig:cost-cell,fig:cc-trace}. From \cref{fig:jsd}, we observed a small $\expval{D_{JS}}$ between the measured bundle traces and their corresponding simulations, indicating little information loss. This suggests that the model effectively captures both local and global features of the measurements, maintaining strong predictive power for a stochastic system.

    \subsubsection{Matrix of Divergences}
    We examined intercellular variation among the measured traces, using the divergence matrix described in \cref{sec:jsd-matrix}. For a given set of traces $\vec{x}(t)$, the divergence matrix reveals this variation. To quantify this variation, we calculated the mean $\expval{D_{JS}}_{i\neq j}$ across the matrix, excluding diagonal elements. Conceptually, this mean represents the variation in motion across different traces. We note that a small $\expval{D_{JS}}$ indicates similar motion, whereas a large value reflects dissimilar motion. The measured-measured divergence matrix yielded a small $\expval{D_{JS}}_{i\neq j}=\num{0.064\pm0.088}$, highlighting low intercellular variation compared to arbitrary functions.
    
    The measured-measured divergence matrix may distinguish between different modalities of bundle movement such as regular \cite{martinSpontaneousOscillationHair2003}, bursting \cite{roongthumskulMultipletimescaleDynamicsUnderlying2011,shlomovitzLowFrequencyEntrainment2013,meenderinkVoltagemediatedControlSpontaneous2015}, and spiking \cite{rutherfordSpikesMembranePotential2009,shlomovitzPhaselockedSpikingInner2014} dynamics. Well-defined clusters within this matrix may correspond to discrete mechanical modes driven by distinct biophysical processes. Conversely, the absence of clusters fails to rule out the existence of distinguishable modes. For example, spiking may represent a neighboring state to bursting, where bursts continuously shorten until they resemble spikes. Thus, hair bundles may exhibit a continuum of mechanical behaviors, transitioning smoothly between the three proposed modes.

    \subsection{Evaluation of the Hair-Bundle Parameters} \label{sec:bundle-params}
    We compared best-fit parameter sets for saccular and AP hair bundles, measuring nine bundles from the sacculus and ten from the AP. We focus our analysis primarily on clusters formed by joint distributions of two parameters, shown in \cref{fig:fit-parameters}. These clusters display comparable centers and shapes across the two end organs. However, the parameter sets for AP bundles appear more dispersed than those for saccular bundles.
    
    Parameters for the AP reveal wider visual spread than those for the sacculus. Potentially, this spread occurs because the AP exhibits richer oscillatory dynamics and poses greater recording challenges. As we reported previously, AP bundles exhibit a broader range of dynamical behaviors, including spiking and bursting oscillations, unlike the typically regular oscillations observed in saccular bundles \cite{vaidoSpontaneousOscillationsHair2025}. Additionally, AP bundles tend to oscillate faster, so individual cycles consist of fewer sample points than the sacculus, which increases fitting error. Moreover, they appear more opaque, decreasing their signal-to-noise ratio and further compromising the parameter fit.

    We examined the best-fit noise strengths acting on the hair bundle and myosin motors. To model these, we added Brownian motions to the differential equations for both components, as detailed in \cref{app:model}. The fits typically yielded stronger noise for the motors than for the bundle, suggesting that thermal fluctuations disproportionately affect the motors. These fit strengths provide recommended parameter ranges, constraining the corresponding physical observables described in \cref{eq:eta-rescale}.

    After adding these noise terms, we toggled them in the model to understand their effects on bundle dynamics. For most bundles, the best-fit simulation oscillated independently of both noise strengths, remaining perpetually either oscillatory or quiescent. Using the best-fit parameters from the stochastic model, we simulated nine non-oscillating bundles in the deterministic model (i.e., with $\nondim{\eta}_{hb}=\nondim{\eta}_a=0$). When introducing bundle noise alone to these bundles (i.e., $\nondim{\eta}_{hb}>0$), we failed to induce an oscillation ($n=9$). However, when adding motor noise (i.e., $\nondim{\eta}_a>0$), we sometimes induced oscillations ($n\leq5$), regardless of noise strength for the bundle. Hence, motor noise partially determines the dynamic state of the hair bundle.

    Motor noise appears essential for observing bursting behavior in hair bundles. Bursting behavior consists of intertwined intervals of quiescent and oscillatory behavior. Simulations of two bursting bundles, based on their best-fit parameters, required motor noise independent of bundle noise. These simulations also revealed a low $\nondim{\tau}_{hb}\approx0.03$ and a high $\nondim{\eta}_{hb}\approx0.08$ relative to the general distribution. By fitting and simulating additional bursting bundles, future studies can identify their underlying biophysical mechanisms.
    
    By examining how simulated parameters shape model behavior, we can explore which internal processes drive bundle dynamics, such as bursting. While many internal cellular processes remain experimentally inaccessible, we can fit their associated parameters to infer how they influence over hair-bundle oscillations. Using our fitting methodology, we can elucidate the underlying dynamics of biophysical systems such as hair bundles.

    \section*{Acknowledgments}
    We thank Justin Faber and Charles Metzler-Winslow for discussions on developing the cost function and cross-correlation distribution. We also acknowledge Justin Faber and Sebastiaan Meenderink for supplying measured traces of spontaneous hair-bundle motion in the sacculus. This work was funded by DoD-AFOSR, Grant FA9550-23-1-0713.

    \appendix

    \section{System-Specific Details for Experimental Methods} \label{app:methods}
    \subsection{Biological Preparation}
    North American bullfrogs (\textit{Rana catesbeiana}) of either gender were anesthetized (pentobarbital: $\SI{150}{\milli\liter\per\kilo\gram}$), double pithed, and decapitated following protocols approved by the University of California, Los Angeles Chancellor’s Animal Research Committee. We extracted sacculi from the inner ears of the frogs and placed them in oxygenated, artificial-perilymph solution (in $\si{\milli\molar}$ as follows: 110 Na$^+$, 2 K$^+$, 1.5 Ca$^{2+}$, 113 Cl$^-$, 3 D-($+$)-glucose, 1 Na$^+$ pyruvate, 1 creatine, 5 HEPES). To mimic the \textit{in vivo} physiological conditions of the surrounding fluid, we mounted the epithelium in a two-compartment chamber. We bathed the apical surfaces in artificial endolymph (in $\si{\milli\molar}$ as follows: 2 Na$^+$, 118 K$^+$, 0.25 Ca$^{2+}$, 118 Cl$^-$, 3 D-($+$)-glucose, 5 HEPES) and basolateral membranes in perilymph \cite{martinSpontaneousOscillationHair2003}. To access individual hair bundles, we removed the otolithic membrane after dissociating it with $\SI{15}{\gram\per\milli\liter}$ of collagenase IV (Sigma-Aldrich) for $\SI{8}{\minute}$.

    Using a similar procedure, we extracted the amphibian papilla---the main auditory organ of the frog---from the dissected inner ears \cite{vaidoSpontaneousOscillationsHair2025}. We mounted the epithelium in a one-compartment chamber, submerging it in an artificial-perilymph solution with lowered calcium ($\SI{0.25}{\milli\molar}$). Analogous to the saccular dissection, we removed the tectorial membrane after dissociating it in collagenase IV for $\SI{9}{\minute}$.

    \subsection{Optical Recordings}
    We collected recordings using an upright optical microscope (Olympus BX51WI) with a water-immersion objective lens (Olympus LUMPlanFL N 60X, NA:1.00), mounted on an optical table (Technical Manufacturing). The microscope was placed inside an acoustically isolated chamber (Industrial Acoustics), minimizing mechanical disturbances to the sensitive hair cells. We recorded innate hair-bundle movement, obtaining 16-bit TIFF images at a resolution of $\SI{108.3\pm0.8}{\nano\meter\per\pixel}$, with a high-speed camera (ORCA-Flash4.0 CMOS) at either 500 or 1000 frames per second. Throughout experiments, the hair bundles sustainably oscillated.
    
    To maximize resolution, we collected differential interference contrast images of the oscillating bundles. Each image accounted for $\SI{1}{\milli\second}$ or $\SI{2}{\milli\second}$ of exposure time, depending on frame rate. To analyze these images, we used custom MATLAB scripts. Specifically, for each frame, we determined the centroid of the bundle weighted by pixel intensity.
    
    To track hair-bundle motion, we calculated position for each video frame. Typical traces yielded noise floors around $\qtyrange{3}{5}{\nano\meter}$, manifesting primarily as thermal \cite{holtonTransductionChannelHair1986,jiAmphibianSacculusForced2018} and ionic \cite{denkForwardReverseTransduction1992,nadrowskiActiveHairbundleMotility2004,barralFrictionTransductionChannels2018} fluctuations. When recording, the hair cell gradually relaxes, slowly drifting the apparent bundle positions. To remove this drift, we subtracted a wide-size, Hann moving average from its corresponding trace.

    \section{Analytic Derivation of Stochastic Terms} \label{app:model}
    To derive the stochastic terms for our model of hair-bundle mechanics, we adapted the dimensional system of ODEs presented in \cite{marcinikComparingParameterreductionMethods2024}. We denote differentials of Brownian motion as $\dd{B}_t\sim\mathcal{N}\qty(0,\dd{t})$, where $\mathcal{N}\qty(\mu,\sigma^2)$ represents a normal distribution with mean $\mu$ and variance $\sigma^2$.
    
    We added stochastic terms to the differentials governing hair-bundle and adaptation-motor positions. We introduced thermal noise to equation (S21), following prior literature on stochastic systems \cite{nadrowskiActiveHairbundleMotility2004,balakrishnanElementsNonequilibriumStatistical2021} to obtain
    \begin{equation}
        \begin{split}
            \lambda_{hb} \dd{X_{hb}} &=
            -\qty[\gamma Nf_{gs} + k_{sp}\qty(X_{hb}-X_{sp})] \dd{t} \\
            &+ \sqrt{2k_BT\lambda_{hb}} \dd{B_{t,hb}} \\
            \dd{x_a} &=
            \qty[S\qty(f_{gs} - k_{es}\qty(x_a+x_{es})) - C] \dd{t} \\
            &+ \sqrt{\frac{2k_BT}{\lambda_a}} \dd{B_{t,a}}
        \end{split}
    \end{equation}
    where $B_{t,hb}$ and $B_{t,a}$ represent independent Brownian motions, such that $\dd{B}_t\sim\mathcal{N}\qty(0,\dd{t})$, for bundle and motor positions, respectively. Consistent with frog saccular hair bundles \cite{denkForwardReverseTransduction1992,vannettenChannelGatingForces2003}, we treat the transduction open-channel probability as deterministic, ignoring channel noise \cite{hanSpontaneousOscillationsSignal2010}.
    
    After nondimensionalizing, the equations take the form
    \begin{equation}
        \begin{split}
            \nondim{\tau}_{hb} \dd{\nondim{x}_{hb}} &= -\qty(\nondim{F}_{gs}-\nondim{x}_{hb})\dd{\nondim{t}} 
            + \nondim{\eta}_{hb} \dd{\nondim{B}_{\nondim{t},hb}} \\
            \nondim{\tau}_a \dd{\nondim{x}_a} &=
            \qty[ \nondim{S}_{max}\nondim{S}\qty(\nondim{F}_{gs}-\nondim{x}_a) - \nondim{C}_{max}\nondim{C}] \dd{\nondim{t}}
            + \nondim{\eta}_a \dd{\nondim{B}_{\nondim{t},a}}
        \end{split}
    \end{equation}
    where $\nondim{B}_{\nondim{t},hb}$ and $\nondim{B}_{\nondim{t},a}$ represent nondimensional Brownian motions, such that $\dd{\nondim{B}}_{\nondim{t}}\sim\mathcal{N}\qty(0,\dd{\nondim{t}})$. The nondimensional noise strengths $\nondim{\eta}_{hb}$ and $\nondim{\eta}_a$ for bundle and motor noise, respectively, relate to their dimensional counterparts by
    \begin{align} \label{eq:eta-rescale}
        \begin{split}
            \nondim{\eta}_{hb} &\coloneq \frac{\chr{\eta}_{hb}}{\sqrt{\chr{\tau}}} \\
            \nondim{\eta}_a &\coloneq \frac{\chr{\eta}_a}{\sqrt{\chr{\tau}}}
        \end{split}
        &
        \begin{split}
            \chr{\eta}_{hb} &\coloneq \frac{\sqrt{2k_BT\lambda_{hb}}}{\gamma Ndk_{gs,max}} \\
            \chr{\eta}_a &\coloneq \frac{\sqrt{2k_BT/\lambda_a}}{C_{max}+S_{max}k_{gs,max}d}
        \end{split}
    \end{align}
    where $\chr{\eta}_{hb}$ and $\chr{\eta}_a$ represent characteristic noise strengths for bundle and motor noise, respectively.
    
    To solve these SDEs numerically, we employed the Euler-Maruyama algorithm \cite{maruyamaContinuousMarkovProcesses1955,sarkkaAppliedStochasticDifferential2019}, adapted from \texttt{sdeint.integrate.itoEuler}. We simulated the system up to nondimensional time $\nondim{t}=500$ using time steps of $\Delta\nondim{t}=2\cdot10^{-3}$.

    \section{Information-Theoretic Measures for Probability Densities} \label{app:measure-compare}
\begin{figure}
        \centering
        \includegraphics{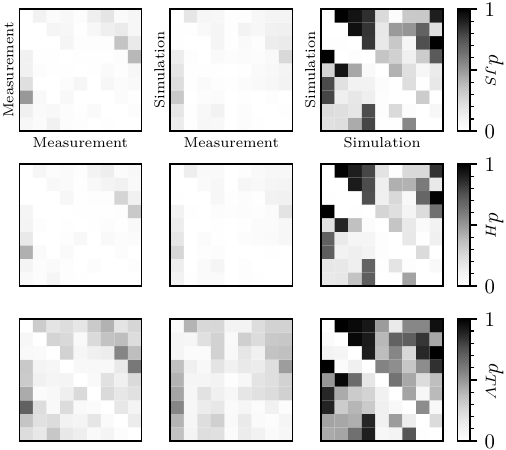}
        \caption{A comparison between alternative divergence matrices, as described in \cref{sec:jsd-matrix}, for nine saccular bundles.
        of alternative information-theoretic measures for divergence matrix.
        Each row indicates a different $L^1$-normalized information-theoretic measure for dissimilarity, corresponding to Jensen-Shannon divergence (top), Hellinger distance (center), and total variation distance (bottom). Each column indicates a different version of divergence matrix, corresponding to measured-measured (left), measured-simulated (center), and simulated-simulated (right).} \label{fig:divergence-compare}
    \end{figure}

    We compared three information-theoretic measures of \enquote{distance} between two probability distributions \cite{chaComprehensiveSurveyDistance2007,sasonDivergenceInequalities2016}, namely total variation distance, Hellinger distance $d_H$, and Jensen-Shannon divergence (JSD), all shown in \cref{fig:divergence-compare}. We chose these three measures specifically because all three remain $L^\infty$-normalized, rendering comparison across datasets feasible. From these three, we chose JSD for further analysis, as it most directly quantifies information loss between a pair of distributions.

    \subsection{Total Variation Distance}
    \begin{figure}
        \centering
        \includegraphics[width=3.375in]{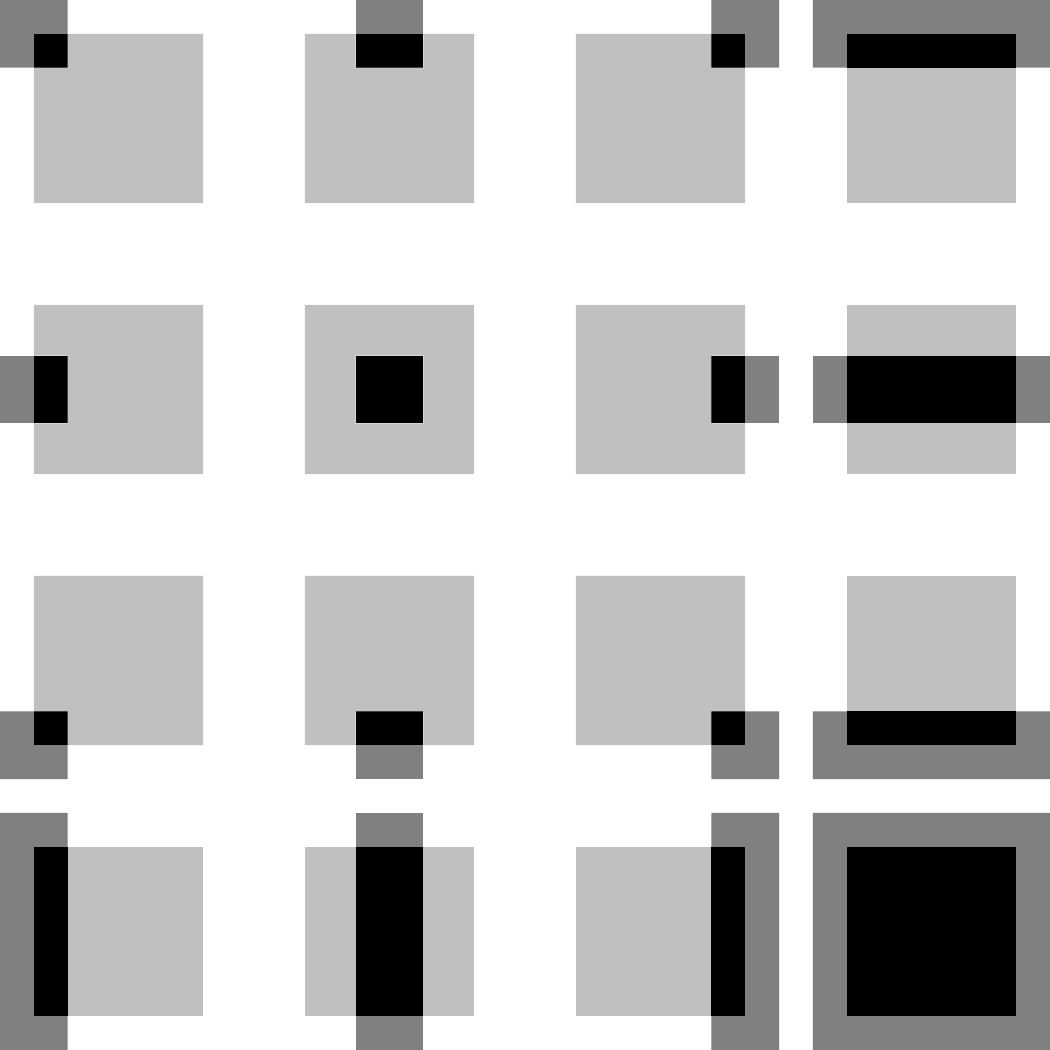}
        \caption{Sixteen possible non-disjoint arrangements for a pair of 2D probability distributions. The (light and dark) gray regions show bounding boxes for the disjoint domains between the two distributions. The black regions show minimal bounding boxes for the intersections of the two distributions.} \label{fig:tvd-arrange}
    \end{figure}
    
    We derived an efficient computational method to calculate TVD. Let $\rho_1(x,y)$ and $\rho_2(x,y)$ represent two joint PDFs. We denote the minimum and maximum $x$ such that $\rho_i(x,y)>0$ as $x_{i,L}$ and $x_{i,R}$, respectively. Similarly, we denote the minimum and maximum $y$ such that $\rho_i(x,y)>0$ as $y_{i,B}$ and $y_{i,T}$, respectively. These bounds yield eight total quantities: minima and maxima $x$ and $y$ for each PDF ($\rho_1(x,y)$ and $\rho_2(x,y)$).
    
    To determine the overlapping bounding box between $\rho_1(x,y)$ and $\rho_2(x,y)$, we found
    \begin{equation}
        \begin{split}
            x_L &= \max\qty{x_{1L}, x_{2L}} \\
            x_R &= \min\qty{x_{1R}, x_{2R}} \\
            y_B &= \max\qty{y_{1B}, y_{2B}} \\
            y_T &= \min\qty{y_{1T}, y_{2T}} \\
        \end{split}
    \end{equation}
    This bounding box defines the smallest region covering all overlaps of $\rho_1(x,y)$ and $\rho_2(x,y)$, as visualized in \cref{fig:tvd-arrange}.
    
    We denote the cumulative distribution function for $\rho_i(x,y)$ as $F_i(x,y)$. We determined the non-overlapping regions of $\rho_i$ toward TVD as
    \begin{equation}
        \begin{split}
            2d_{i,TV} &= 1 - F_i(x_R,y_T) \\
            &+ F_i(x_R,y_B) + F_i(x_L,y_T) - F_i(x_L,y_B)
        \end{split}
    \end{equation}
    where $\abs{\cdot}$ represents the absolute value or complex modulus. We then calculated the overall TVD between $\rho_1(x,y)$ and $\rho_2(x,y)$ as
    \begin{equation} \label{eq:tvd-fast}
        \begin{split}
            d_\text{TV} \qty(\rho_1, \rho_2) &= d_{1,TV} + d_{2,TV} \\
            &+ \frac{1}{2} \int_{y_B}^{y_T}\int_{x_L}^{x_R} \abs{\rho_2(x,y)-\rho_1(x,y)} \dd{x}\dd{y}
        \end{split}
    \end{equation}
    Conceptually, the terms $d_{1,TV}$ and $d_{2,TV}$ account for the non-overlapping regions of $\rho_1$ and $\rho_2$, whereas the integral accounts for the overlapping region. This hybrid formulation improves computational speed and numerical precision.

    \subsection{Hellinger Distance and Jensen-Shannon Divergence}
    We calculated $d_H$ and JSD by expressing them as their respective $f$-divergences \cite{qiaoFdivergenceGeneralizedInvariant2008,nielsenJensenShannonSymmetrization2019}. For $d_H$, we computed only the overlapping region between $\rho_1$ and $\rho_2$, as non-overlapping regions leave the total measure unchanged. Across all measures of dissimilarity, we used continuous PDFs rather than discrete distributions to ensure these comparisons remained $L^1$-normalized.
    
    \section{Computational Details for Fitting Procedure} \label{app:cost-function}        
    \subsection{Application of Differential Evolution} \label{app:de}
    We used \texttt{scipy.optimize.differential\_evolution} to perform differential evolution \cite{stornDifferentialEvolutionSimple1997,priceDifferentialEvolution2005}. For our differential evolution hyperparameters, we set population size to 64 \cite{piotrowskiReviewDifferentialEvolution2017}, initial guesses to Sobol sampling \cite{renardySobolNotSobol2021}, strategy to \enquote{rand1exp} \cite{qiangUnifiedDifferentialEvolution2014}, recombination to 0.7, and mutation to dynamically between 0.5 and 1. Notably, we changed the initial guesses, population size, and strategy from the defaults in \texttt{scipy.optimize.differential\_evolution}.
    
    During differential evolution, we declared the same noise seed before each simulation and across iterations. As a result, we asserted the same noise instances independently for $\dd\nondim{B}_{\nondim{t},hb}$ and $\dd\nondim{B}_{\nondim{t},a}$, only multiplying them by their respective noise strengths $\nondim{\eta}_{hb}$ and $\nondim{\eta}_{a}$.

    \subsection{Components of Cost Function}
    \begin{figure}
        \centering
        \includegraphics{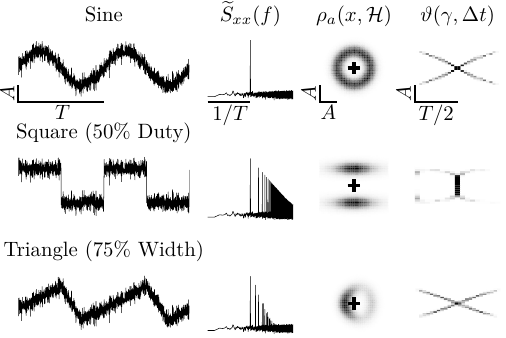}
        \caption{The three cost-function components, described in \cref{sec:cost}, for three common waveforms. Rows correspond to waveforms of sine (top), square (50\% duty, center), and triangle (75\% width, bottom). Column 1: The solid lines display position $x$ over time $t$, each with zero mean position. The scale bar displays amplitude $A$ along the $y$-axis and period $T$ along the $x$-axis. Column 2: In a log-log scale, the solid lines display power spectral density $\widetilde{S}_{xx}$ over frequency $f$, and the dashed lines indicate their median frequency. The scale bar displays frequency $1/T$ along the $x$-axis. Column 3: The heatmaps display PDF $\rho_a(x,\mathcal{H})$ over position $x$ along the $y$-axis and Hilbert transform $\mathcal{H}\qty{x}$ along the $x$-axis. The plus signs indicate the origin of $(x,\mathcal{H}\qty{x})$. The scale bar displays amplitude along both axes. Column 4: The heatmaps display PDF $\vartheta\{x\}(\gamma,\Delta{t})$ over crossing position $\gamma$ along the $y$-axis and time difference $\Delta{t}$ along the $x$-axis. The scale bar displays amplitude along the $y$-axis and half-period along the $x$-axis.} \label{fig:cost-demo}
    \end{figure}

    As components of the cost function, we evaluated the power spectral density (PSD), the distribution of analytic signal (DAS), and the distribution of position crossings (DPC) to characterize oscillations. For each of these representative functions, we used TVD as a measure of dissimilarity. While statisticians traditionally apply TVD to compare two probability distributions, we applied it more broadly to $L^1$-normalized functions.

    \subsubsection{Shape of Power Spectral Density} \label{app:psd}
    The PSD contributed 10\% to the total weight of our cost function, capturing primarily the shape and timescale of a trace. We denote the PSD of position $x$ as $S_{xx}\qty{x}(f)$. To effectively smooth power spectral density, we used the Bartlett method \cite{bartlettPeriodogramAnalysisContinuous1950,welchUseFastFourier1967} to segment the trace into eight non-overlapping parts.
    
    We $L^1$-normalized this PSD as $\widetilde{S}_{xx}\qty{x}(f)$. This normalized version resembles a PDF with unit area. Thus, we quantified the dissimilarity between measured and simulated PSDs by evaluating their TVD.
    
    To convert from simulated into measured position, we directly rescaled
    \begin{equation}
        \begin{split}
            \widetilde{S}_{xx} &\mapsto \chr{\tau} \widetilde{S}_{xx} \\
            f &\mapsto \frac{f}{\chr{\tau}}
        \end{split}
    \end{equation}
    This rescaling relies solely on $\chr{\tau}$, remaining independent of $\chr{x}_{hb}$ and $\check{x}_{hb}$, which implies that the normalized PSD efficiently rescales time in particular.

    \subsubsection{Distribution of Analytic Signal} \label{app:das}
    The DAS contributed 50\% to the total weight of our cost function, capturing both the shape and amplitude of a trace. We denote the joint PDF of the analytic signal over position $x$ and its Hilbert transform $\mathcal{H}\qty{x})$ as $\rho_a(x,\mathcal{H}\qty{x})$. To quantify the dissimilarity between measured and simulated DASs, we evaluated their TVD.
    
    To convert from simulated into measured position, we rescaled
    \begin{equation}
        \begin{split}
            \rho_a &\mapsto \frac{\rho_a}{\chr{x}_{hb}^2} \\
            x &\mapsto \chr{x}_{hb} \qty(x - \check{x}_{hb}) \\
            \mathcal{H}\qty{x} &\mapsto \check{x}_{hb} \mathcal{H}\qty{x}
        \end{split}
    \end{equation}
    leveraging the identity $\mathcal{H}\qty{ax+b}=a\mathcal{H}\qty{x}$ with constants $a,b$ \cite{hahnComplexHypercomplexAnalytic2017}. This rescaling depends exclusively on $\chr{x}_{hb}$ and $\check{x}_{hb}$, not on $\chr{\tau}$, which suggests that the nondimensional DAS methodically rescales position.
    
    We determined the number of bins for each axis of the analytic signal dynamically. Let $N_x$ denote the number of points in the computational array for trace $x(t)$. We set the number of bins for $x$ and $\mathcal{H}\qty{x}$ individually to $\floor{\sqrt[3]{N_x}}$, where $\floor{\cdot}$ represents the rounding function (to the nearest integer). This cube-root heuristic balances the number of bins along each $x$ and $\mathcal{H}\qty{x}$ as well as points per bin.

    \subsubsection{Distribution of Position Crossings} \label{app:dpc}
    The DPC contributed 40\% of the total weight to our cost function, capturing the shape, timescale, and amplitude of a trace. We denote the joint PDF of position crossings over crossing position $\gamma$ and half-period $\Delta{t}$ as $\vartheta\qty(\gamma,\Delta{t})$. Before calculating this distribution, we denoised traces using wavelets \cite{changAdaptiveWaveletThresholding2000}. Computationally, we chose the \enquote{VisuShrink} method \cite{donohoIdealSpatialAdaptation1994} with the \enquote{soft} mode and \enquote{sym4} wavelet \cite{arfaouiWaveletAnalysisBasic2021} via \texttt{skimage.restoration.denoise\_wavelet}. To quantify the dissimilarity between measured and simulated DPCs, we evaluated their TVD.
    
    To convert from simulated to measured position, we rescaled
    \begin{equation}
        \begin{split}
            \vartheta &\mapsto \frac{\vartheta}{\chr{\tau}\chr{x}_{hb}} \\
            \gamma &\mapsto \chr{x}_{hb} \qty(\gamma - \check{x}_{hb}) \\
            \Delta{t} &\mapsto \chr{\tau} \Delta{t}
        \end{split}
    \end{equation}
    This rescaling depends on all three variables (i.e., $\chr{x}_{hb}$, $\check{x}_{hb}$, $\chr{\tau}$), meaning the nondimensional DPC versatilely rescales the simulated traces in full.
    
    We included only robust crossing events, ignoring extraneous events due to noise. A trace $x$ can register multiple crossings during a single event, by fluctuating stochastically through a position multiple times in quick succession. To mitigate unwarranted crossings from noise, we used bins surrounding $\gamma$ rather than individual values.

    We assigned a single representative time to each $\gamma$-crossing event. Initially, we characterized each event computationally with multiple crossing times as the trace passed through the boundaries of a bin. These degenerate crossing times occurred because the trace crossed through both boundaries, potentially multiple times on either one. To consolidate these into a single crossing time, we calculated the midpoint between the final two crossing times, one through each boundary of the bin.
    
    To generate the tuple $\zc[\gamma]{x}$ of $\gamma$-crossing times, we analyzed the denoised position $x(t)$. We denote the ordered tuples of upward- and downward-crossing times as $\zc[\gamma\uparrow]{x}$ and $\zc[\gamma\downarrow]{x}$, respectively, both sorted in ascending order of time. Furthermore, we represent the number of crossings in these tuples as $N_{\gamma\uparrow}\coloneq\abs{\zc[\gamma\uparrow]{x}}$ and $N_{\gamma\downarrow}\coloneq\abs{\zc[\gamma\downarrow]{x}}$, respectively. 
    
    We counted a $\gamma$-crossing only when $x$ crossed fully through a bin around $\gamma$. For a small neighborhood size $\delta{x}>0$, we separately counted upward and downward crossings. We counted an upward crossing in $\zc[\gamma\uparrow]{x}$, incrementing $N_{\gamma\uparrow}$, when $x$ crossed consecutively through $\gamma-\delta{x}$ followed by $\gamma+\delta{x}$. Similarly, we counted a downward crossing in $\zc[\gamma\downarrow]{x}$, incrementing $N_{\gamma\downarrow}$, when $x$ crossed consecutively through $\gamma+\delta{x}$ followed by $\gamma-\delta{x}$.
    
    We determined a number $N_\gamma$ of equidistant bin edges for crossing position $\gamma$ manually. In general, the optimal number for $N_\gamma$ depends on the signal-to-noise ratio of the position \cite{tanweerRobustCrossingsDetection2024}, which varies across biophysical systems. Smaller values for $N_\gamma$ provide greater precision for $\vartheta\qty(\gamma,\Delta{t})$, whereas larger values yield greater precision for $\gamma$. To balance these trade-offs, we chose $N_\gamma=21$ arbitrarily after analyzing several hair-bundle traces, achieving sufficient precision in both $\vartheta\qty(\gamma,\Delta{t})$ with $\gamma$. We selected the minimum and maximum values for $\gamma$ to span those in $x$.
    
    After discretizing bin edges for $\gamma$, we determined a number $N_{\Delta{t}}$ of equidistant bin edges for half-period $\Delta{t}$ dynamically. Specifically, we calculated
    \begin{equation}
        \begin{split}
            N_{\Delta{t}} &= \left\lfloor \sqrt[6]{N_\uparrow N_\downarrow} \right\rfloor \\
            N_{(\uparrow\downarrow)} &\coloneq \sum_\gamma N_{\gamma(\uparrow\downarrow)}
        \end{split}
    \end{equation}
    where $N_\uparrow$ and $N_\downarrow$ represent the total number of upward- and downward-crossing times, respectively. Conceptually, this formula computes the cube root of the geometric mean of $N_\uparrow$ and $N_\downarrow$. Similar to the heuristic in \cref{app:das}, we included this cube root to balance the number of bins along each $\gamma$ and $\Delta{t}$ with points per bin.
    
    Finally, we selected the minimum $\Delta{t}_{min}$ and maximum $\Delta{t}_{max}$ bin-edge values for $\Delta{t}$ to balance coverage and accuracy. Let $\zc[]{x}\coloneq\bigcup_\gamma\zc[\gamma]{x}$ represent the ordered tuple containing all crossing times in $\zc[\gamma]{x}$ across all $N_\gamma$ values of $\gamma$, sorted in ascending order of time. We chose
    \begin{equation}
        \begin{split}
            \Delta{t}_{min} &= \min\zc[]{x} \\
            \Delta{t}_{max} &= \sqrt{N_\gamma} \max\zc[]{x}
        \end{split}
    \end{equation}
    By incorporating the extent of $\zc[]{x}$, we covered a broad range over $\Delta{t}$. By including the scaling factor $\sqrt{N_\gamma}$, we compensated for the shorter crossing times in $\zc[]{x}$ compared to those in individual $\zc[\gamma]{x}$ tuples.

    \subsubsection{Velocity Field of Analytic Signal} \label{app:velocity-field}
    As an alternative cost-function component, we examined the 2D velocity field $\dnasig{x}$ of trace $x(t)$. Using \texttt{scipy.stats.binned\_statistic\_2d}, we binned this field along $(x,\mathcal{H}\qty{x})$ to generate a 2D joint PDF, analogous to $\rho_a\qty{x}$ in \cref{eq:asigp}. To quantify the discrepancy between velocity fields of the measured and simulated traces, we computed their TVD, taking the complex modulus in \cref{eq:tvd-fast}. Ultimately, we excluded this component as part of the cost function because this PDF remained too sensitive to transient noise in the traces.

    \subsection{Validation on Triangle Waves} \label{app:triangle-recovery}
    \begin{figure}
        \centering
        \includegraphics{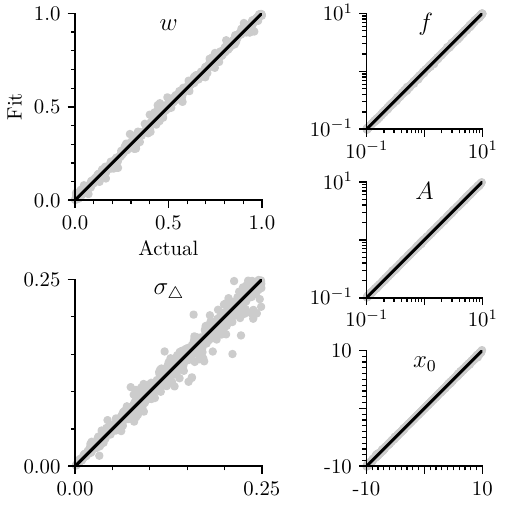}
        \caption{A comparison between fit and actual parameters for simulated triangle waves. The plot titles indicate the parameter for the triangle wave in \cref{eq:triangle}: width $w$, noise strength $\eta_\triangle$, frequency $f$, amplitude $A$, and offset $x_0$. The $y$-axis shows fit parameter values, and the $x$-axis shows actual parameter values. The markers show pairs of fit and actual parameter values for simulated triangle waves. The lines indicate where fit and actual values equal each other.} \label{fig:triangle-recovery}
    \end{figure}
    
    To validate the cost function, we recovered parameter values for 600 simulated triangle waves using \cref{eq:triangle}. For each wave, we generated five parameters using \texttt{numpy.random.uniform} as follows:
    \begin{equation}
        \begin{split}
            w &\in U[0, 1) \\
            \sigma_\triangle &\in U[0, 0.25) \\
            f, A &\in 10 ^ {U[-1, 1)} \\
            x_0 &\in U[-10, 10)
        \end{split}
    \end{equation}
    where $U[a,b)$ represents a uniform random variable spanning the interval $[a,b)$.
    
    Computationally, we first calculated the deterministic triangle wave. During each simulation, we covered a time interval $t\in[0, 20]$ with a fixed time step of $\Delta{t}=2\cdot10^{-3}$, generating an array with 20000 points. We then added noise independently to each point, sampling from $\mathcal{N}(0,1)$ and scaling by noise strength. In \cref{fig:cost-demo}, we exemplified a triangle wave with $w=0.75$ and $\sigma_\triangle=0.25$.
    
    For each simulated triangle wave, we fit parameters using the cost function described in \cref{sec:cost}. During differential evolution, we reused a single instance of noise, scaling by the guessed noise strength in each parameter set. By fixing the noise seed, we mitigated the risk of anchoring prematurely to an unstable minimum, caused by noise realizations with unusually low cost.

    \subsection{Cross-Correlation of Trace Segments}
    \begin{figure}
        \centering
        \includegraphics{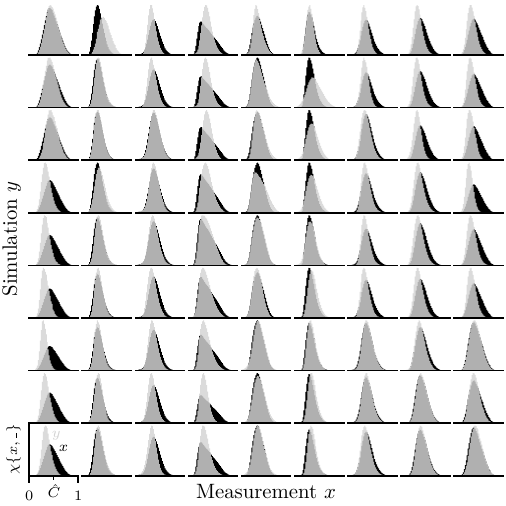}
        \caption{A comparison between measured and simulated distributions of cross-correlations for nine saccular bundles. Columns (large $x$-axis) correspond to measured bundle traces $x(t)$, and rows (large $y$-axis) correspond to simulated bundle traces $y(t)$. The small $x$-axis displays cross-correlation $\hat{C}$. The small $y$-axis displays PDF $\chi(\hat{C})$, ranging up to the maximum probability density per plot. Black histograms show PDF $\chi\qty{x,x}(\hat{C})$, which correlates a measured bundle to itself. Gray histograms show PDF $\chi\qty{x,y}(\hat{C})$, which correlates a measured bundle to a simulated bundle.} \label{fig:cc-distribution}
    \end{figure}
    
    Segment duration affects the divergence matrices described in \cref{sec:jsd-matrix}, so we calculated these matrices for $T=4,8,12,16,20$ cycles. Ultimately, we chose segment duration $T$ equal to four periods of reference trace $x(t)$, calculating periods as the reciprocal median frequency in $S_{xx}\qty{x}$. Relative to larger durations, this choice widened the cross-correlation distributions, yielding a median around 0.5 as shown in \cref{fig:cc-distribution}.
    
    We ranked values in the measured-simulated divergence matrices. We then performed Page's trend test ($p<0.001$; \cite{pageOrderedHypothesesMultiple1963}), which indicated consistent rankings in the divergence matrices across different durations. Furthermore, we performed linear regression pairwise on these divergence matrices, resulting in $r^2=\num{0.977\pm0.026}$, indicating that JSD increased by a roughly constant factor with changing $T$. When performing the same test on the other divergence matrices (e.g., measured-measured, simulated-simulated), these rankings retained low $p$-value and high $r^2$-value.

    We performed linear regression to recover the divergence matrices at different cycle lengths. Letting $\jsd{x_T}{y_T}$ represent divergence from the matrix corresponding to segment duration $T$ cycles, we fit the curve
    \begin{equation}
        \frac {\jsd{x_T}{y_T}} {\jsd{x_4}{y_4}} = a (1 + bT)
    \end{equation}
    where $a,b\in\mathbb{R}$, a form justified by the high $r^2$ between $\jsd{x_4}{y_4}$ and $\jsd{x_T}{y_T}$ for $T=4,8,12,16,20$. Using least-squares regression, we found $a=\num{0.815\pm0.035}$ and $b=\num{0.035\pm0.006}$. These values for $a,b$ can be used to approximate $\jsd{x_T}{y_T}$ at different segment durations. This linear regression fails for large $T$ because $\jsd{x_T}{y_T}$ asymptotically approaches a constant value as $T\rightarrow\infty$. However, this regression remains accurate for $T$ between 4 and 20 cycles.

    \section{Python Packages}
    \texttt{numpy} \cite{harrisArrayProgrammingNumPy2020},
    \texttt{scipy} \cite{virtanenSciPy10Fundamental2020},
    \texttt{sdeint} \cite{aburnCriticalFluctuationsCoupling2017},
    \texttt{skimage} \cite{waltScikitimageImageProcessing2014},
    \texttt{matplotlib} \cite{hunterMatplotlib2DGraphics2007},
    \texttt{pandas} \cite{mckinneyDataStructuresStatistical2010}

    \begin{figure*}
        \centering
        \includegraphics{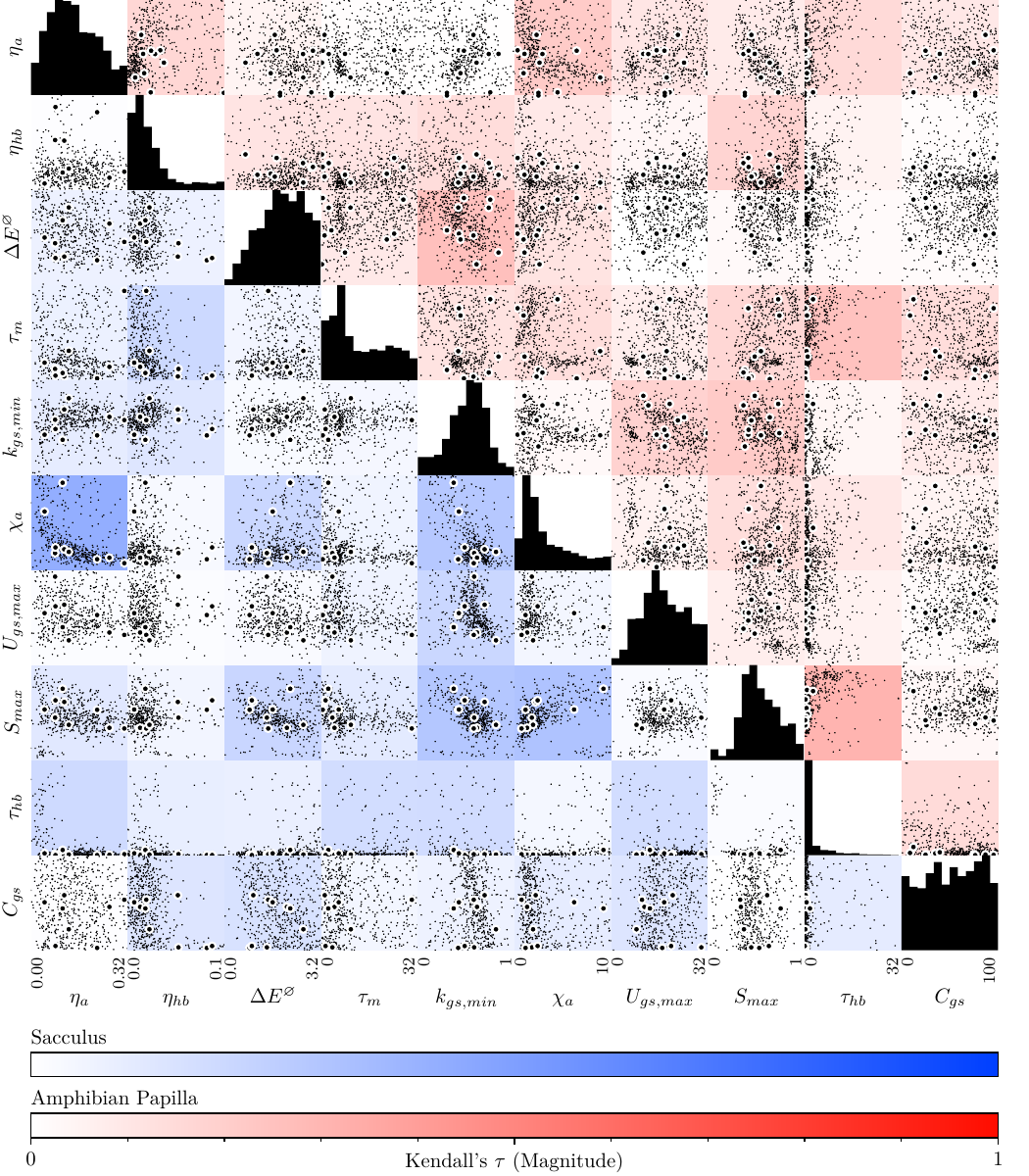}
        \caption{The pairwise distributions of best-fit parameters for sacculi and amphibian papillae. On-diagonal: The histograms show PDFs of all parameters during the final iteration of differential evolution, combining saccular and papillar bundles. Off-diagonal: Blue (bottom left) and red (top right) backgrounds correspond to saccular and AP bundles, respectively. The $x$- and $y$-axes range from minimum to maximum parameter bounds used when fitting. Large markers display best-fit parameters during the final iteration, nine for the sacculus and ten for the papilla. Small markers display all parameters during the final iteration (population size 64), with $9\times64=576$ for the sacculus and $10\times64=640$ for the papilla. The colorbars indicate magnitude of Kendall's $\tau$, displayed as saturation of background color, for each pair of parameters (entire population), distinguishing between the sacculus (blue) and papilla (red).} \label{fig:fit-parameters}
    \end{figure*}

    \clearpage
    \bibliography{reference}

\end{document}